\input harvmac

\def\Title#1#2{\rightline{#1}\ifx\answ\bigans\nopagenumbers\pageno0
\else\pageno1\vskip.5in\fi \centerline{\titlefont #2}\vskip .3in}

\font\caps=cmcsc10

\noblackbox
\parskip=1.5mm



\def\CN{{\cal N}}


\def\dj{\hbox{d\kern-0.347em \vrule width 0.3em height 1.252ex depth
-1.21ex \kern 0.051em}}

\def\Dirac{\,\raise.15ex\hbox{/}\mkern-13.5mu D}
\def\dirac{\,\raise.15ex\hbox{/}\kern-.57em \partial}
\def\shalf{{\ifinner {\scriptstyle {1 \over 2}}\else {1 \over 2} \fi}} 

\lref\malda{
J.~Maldacena,
Adv.\ Theor.\ Math.\ Phys.\  {\bf 2}, 231 (1998)
[hep-th/9711200].
}

\lref\gkpw{
S.~S.~Gubser, I.~R.~Klebanov and A.~M.~Polyakov,
Phys.\ Lett.\  {\bf B428}, 105 (1998)
[hep-th/9802109];
E.~Witten,
Adv.\ Theor.\ Math.\ Phys.\  {\bf 2}, 253 (1998)
[hep-th/9802150].
}

\lref\cds{
A.~Connes, M.~R.~Douglas and A.~Schwarz,
JHEP {\bf 9802}, 003 (1998)
[hep-th/9711162].
}

\lref\dh{
M.~R.~Douglas and C.~Hull,
JHEP {\bf 9802}, 008 (1998)
[hep-th/9711165].
}

\lref\sw{
N.~Seiberg and E.~Witten,
JHEP {\bf 9909}, 032 (1999)
[hep-th/9908142].
}

\lref\hi{
A.~Hashimoto and N.~Itzhaki,
Phys.\ Lett.\  {\bf B465}, 142 (1999)
[hep-th/9907166].
}

\lref\mr{
J.~M.~Maldacena and J.~G.~Russo,
JHEP {\bf 9909}, 025 (1999)
[hep-th/9908134].
}

\lref\ncdp{M.~Li and Y.~Wu,
[hep-th/9909085].
}

\lref\ncyaron{M.~Alishahiha, Y.~Oz and M.~M.~Sheikh-Jabbari,
JHEP {\bf 9911}, 007 (1999)
[hep-th/9909215].
}

\lref\holow{N.~Dorey, T.~J.~Hollowood, V.~V.~Khoze, 
M.~P.~Mattis and S.~Vandoren,
Nucl.\ Phys.\  {\bf B552}, 88 (1999)
[hep-th/9901128].
}

\lref\diracprop{
B.~Allen and C.~A.~Lutken,
Commun.\ Math.\ Phys.\  {\bf 106}, 201 (1986);
T.~Kawano and K.~Okuyama,
[hep-th/9905130].
}

\lref\bprop{
C.~P.~Burgess and C.~A.~Lutken,
Phys.\ Lett.\  {\bf B153}, 137 (1985);
T.~Inami and H.~Ooguri,
Prog.\ Theor.\ Phys.\  {\bf 73}, 1051 (1985).
}

\lref\burg{
C.~J.~Burges, D.~Z.~Freedman, S.~Davis and G.~W.~Gibbons,
Annals Phys.\  {\bf 167}, 285 (1986).
}

\lref\banks{T.~Banks and M.~B.~Green,
JHEP {\bf 9805}, 002 (1998)
[hep-th/9804170];
M.~Bianchi, M.~B.~Green, S.~Kovacs and G.~Rossi,
JHEP {\bf 9808}, 013 (1998)
[hep-th/9807033];
}

\lref\jose{J.~L.~F.~Barbon,
Phys.\ Lett.\  {\bf B404}, 33 (1997)
[hep-th/9701075].
}

\lref\greengut{M.~B.~Green and M.~Gutperle,
Nucl.\ Phys.\  {\bf B498}, 195 (1997)
[hep-th/9701093].
M.~B.~Green,
[hep-th/9903124].
}

\lref\greenal{M.~B.~Green,
Phys.\ Lett.\  {\bf B329}, 435 (1994)
[hep-th/9403040];
M.~B.~Green,
Phys.\ Lett.\  {\bf B354}, 271 (1995)
[hep-th/9504108];
}

\lref\kogan{J.~L.~F.~Barbon, I.~I.~Kogan and E.~Rabinovici,
Nucl.\ Phys.\  {\bf B544}, 104 (1999)
[hep-th/9809033].
}

\lref\topsus{E.~Witten,
Phys.\ Rev.\ Lett.\  {\bf 81}, 2862 (1998)
[hep-th/9807109];
A.~Hashimoto and Y.~Oz,
Nucl.\ Phys.\  {\bf B548}, 167 (1999)
[hep-th/9809106].
}

\lref\nekrasov{N.~Nekrasov and A.~Schwarz,
Commun.\ Math.\ Phys.\  {\bf 198}, 689 (1998)
[hep-th/9802068];
A.~Mikhailov,
[hep-th/9910126];
A.~Dhar, G.~Mandal, S.~R.~Wadia and K.~P.~Yogendran,
[hep-th/9910194];
S.~R.~Das, S.~Kalyana Rama and S.~P.~Trivedi
[hep-th/9911137].
}

\lref\ncthermobr{
J.~L.~F.~Barbon and E.~Rabinovici,
JHEP {\bf 9912}, 017 (1999)
[hep-th/9910019].
}

\lref\ncthermoo{R.~Cai and N.~Ohta,
[hep-th/9910092];
T.~Harmark and N.~A.~Obers,
JHEP {\bf 0001}, 008 (2000)
[hep-th/9910036].
}

\lref\usold{J.~L.~Barbon and A.~Pasquinucci,
Phys.\ Lett.\  {\bf B458}, 288 (1999)
[hep-th/9904190].
}

\lref\kkred{H.~J.~Kim, L.~J.~Romans and P.~van Nieuwenhuizen,
Phys.\ Rev.\  {\bf D32}, 389 (1985);
M.~Gunaydin and N.~Marcus,
Class.\ Quant.\ Grav.\  {\bf 2}, L11 (1985).
}

\lref\marti{T. Banks, M. R. Douglas, G. T. Horowitz and E. Martinec,
[hep-th/9808016]. }


\baselineskip=14pt

\line{\hfill CERN-TH/2000-062}
\line{\hfill IFUM/653}
\line{\hfill Bicocca-FT-99-42}
\line{\hfill {\tt hep-th/0002187}}
\vskip 1.0cm

\Title{\vbox{\baselineskip 12pt\hbox{}
 }}
{\vbox {\centerline{A Note on Interactions of (Non-Commutative)}
\vskip10pt
\centerline{Instantons Via AdS/CFT}
}}

\vskip0.6cm

\centerline{$\quad$ {\caps J. L. F. Barb\'on~\foot{ On leave
from  ``Departamento de F\'{\i}sica de Part\'{\i}culas,
Universidad de Santiago de Compostela, Spain".}
and A. Pasquinucci~$^{2}$
}}
\vskip0.8cm

\centerline{{\sl $^1$ Theory Division, CERN}}
\centerline{{\sl CH-1211 Geneva 23, Switzerland}}
\centerline{{\tt barbon@mail.cern.ch}}

\vskip0.3cm

\centerline{{\sl $^2$ Dipartimento di Fisica, Universit\`a di Milano }}
\centerline{{\sl and Dipartimento di Fisica, Universit\`a Milano-Bicocca }}
\centerline{{\sl and INFN, sezione di Milano}}
\centerline{{\sl via Celoria 16, I-20133 Milano, Italy}}
\centerline{{\tt andrea.pasquinucci@cern.ch}} 

\vskip 0.2in
\vfill

\noindent We consider the interaction between instantons and anti-instantons
in four-dimensional $\CN=4$ super-Yang--Mills theory at large $N$ and 
large 't Hooft coupling as described by D-instantons via AdS/CFT duality. 
We give an estimate of the strength of the interaction in various regimes. 
We discuss also the case of Non-Commutative super-Yang--Mills theory where 
the interaction between instantons and anti-instantons can be used
as a way to probe the locality properties of the theory in
the supergravity picture, without explicit reference to the definition 
of local operators.




\Date{February 2000}



\baselineskip=14pt

\newsec{Introduction and Conclusions}

One of the most succesful tests of the AdS/CFT correspondence
of refs.\ \refs{\malda,\gkpw} is provided by instanton physics.
In supersymmetric theories, many aspects of instanton dynamics
(such as the moduli space and the measure) are of BPS-type, i.e.\   
they are protected and thus provide useful tests of the duality.

The AdS/CFT picture of Yang--Mills instantons is in terms of D$(p-4)$-brane
probes in the background geometry (vacuum) of D$p$-branes \refs{\banks}. 
Concentrating in the case of the $\CN=4$ SYM theory,
dual to type IIB string theory on 
 $AdS_5 \times {\bf S}^5$, we can write the background  metric as
\eqn\adsmet
{ 
{ds^2 \over R^2} = {1\over \rho^2} \left( d\rho^2 + d{\vec x}^{\,2} \right) +
d\Omega_5^2,}
where the radius $R$ is determined by the Regge slope and string
coupling by  $R^4 =4\pi \, \alpha'^2 \,g_s \,N = \alpha'^2 \,\lambda$. We
have also defined the 't Hooft coupling $\lambda = g_s \,N = g_{\rm YM}^2 \,N$
controlling the large-$N$ expansion of the gauge theory. 
 The coordinates ${\vec x}$ parametrize the ${\bf R}^4$ boundary
giving space-time data for the gauge theory, while $\rho$ represents
the gauge-theory length scale according to the UV/IR relation. For example,
a D-instanton probe sitting at $(\rho, {\vec x})$ is interpreted as a
Yang--Mills instanton of size parameter $\rho$
 located at point ${\vec x}$ in ${\bf R}^4$.

The calculations of instanton-dominated BPS
 amplitudes using the AdS/CFT rules
match dramatically with
the perturbative computations recently done at weak coupling \refs{\holow}. 
In this paper instead we are interested in studying non-BPS
configurations containing both instantons  and anti-instantons (thus breaking
all  
supersymmetries) still using semi-classical physics\foot{See \refs{\usold} for
other aspects of non-supersymmetric instanton physics.}. Of course in this
case we do not expect the perturbative results to match the supergravity
ones, which should describe the physics at strong coupling. The results
we obtain are then predictions which could help us in understanding the
strong coupling regime of large-$N$ $\CN=4$ SYM theory.

Since the dual description is
gauge invariant, it is problematic to describe the collective coordinates
associated to relative gauge orientations in a multi-instanton configuration.
Presumably we should interpret the gravitational description as providing the
result of having integrated out the gauge collective coordinates. On the other
hand, the D-instanton geometrical moduli   corresponding to the location in
${\bf S}^5 $ do have gauge-theory   interpretation in terms of new collective
coordinates parametrizing large $N$ saddle-point approximations  of the
instanton measure \refs{\holow}. Therefore, as long as we have the hierarchy
$1\ll \lambda \ll N$ we should be able to study  instanton dynamics
of the SYM theory via dilute D-instantons in the supergravity approximation
to IIB strings on $AdS_5 \times {\bf S}^5$.          

Since the D-instantons are fully localized in ten dimensions, we expect
to find  physical quantities in the four-dimensional SYM theory
 revealing (at least for $\lambda \gg 1$) a ten-dimensional
scaling.  The interaction of I/A pairs is a convenient probe of the localization
properties, in the sense that the interaction will show ten-dimensional
features if the proper distance between the I/A pair is much smaller than
the curvature radius of $AdS_5 \times {\bf S}^5$. For sufficiently large
values of the 't Hooft coupling, this condition is compatible with
the dilute character of the D-instanton gas in supergravity. In fact, we
find that the dilute D-instanton gas in supergravity  corresponds
to the dilute gas of SYM instantons {\it only} 
 in the limit where the I/A distances
are much larger than the curvature radius, i.e. when no ten-dimensional features are revealed. On the other hand, the fully localized regime mentioned above
is to be interpreted as a transient phase of overlapping YM instantons, i.e.
the geometrical
 notions of ``diluteness" are slightly different on both sides
of the AdS/CFT correspondence.

 The occurence in four-dimensional gauge
theories of regimes with ten-dimensional features via the AdS/CFT duality
is not new. For example, a similar phenomenon is found in the study of the
dependence of the entropy on the energy scales of the theory, i.e. there
are intermediate energy regimes, visible at large $\lambda$, which
show ten-dimensional scaling of the entropy \refs{\marti}. One of these
regimes is characterized by fully localized configurations in $AdS_5 \times
{\bf S}^5$ where the density of states is well approximated by that of
ten-dimensional Schwarzschild black holes. We could present our results
as the euclidean, semiclassical analog of these localized states. The
analogy with the entropy estimates is rather close, since we are
also looking at non-BPS quantities and the large-$\lambda$ results
are to be interpreted as predictions of AdS/CFT, rather than tests
to be satisfied.

We shall furthermore compare the
results with the case where we add a non-vanishing Neveu--Schwarz (NS)
$B$-field to the background configuration. 
Indeed, given the embedding of Non-Commutative Yang--Mills theories (NCYM) 
in open string dynamics with non-vanishing
Neveu--Schwarz (NS) $B$-field \refs{\cds,\dh,\sw},
the possibility opens up of studying the large $N$ limit of
these systems by means of generalizations of the AdS/CFT correspondence.
In the recent works \refs{\hi} and \refs{\mr} a gravitational dual of
large $N$ 
NCYM was proposed as the near-horizon geometry of D$p$-branes with
non-trivial profiles of NS $B$-fields, under the 
appropriate scaling of parameters that isolates the low energy regime
with NCYM physics.

We will  consider the interaction action of
an instanton/anti-instanton pair as an intermediate probe into the locality
properties of the NCYM theory. Although the dependence of such
quantity on the pair's spatial separation is providing some local
information, one can compute it using the AdS/CFT rules without
explicitly defining local operators via boundary behaviour, a rather
problematic procedure in view of the results of \refs{\mr}. In particular,
one can define this quantity in terms of the {\it bulk}
D-instanton/anti-D-instanton
interaction action in Type IIB superstring theory                    

We will thus be able to see how the non-locality properties of NCYM are
reflected in a different behaviour of the I/A interactions. We find
evidence of an effective delocalization of instantons of size smaller
than the non-commutativity length scale.

\newsec{The UV/IR relation and I/A interactions}

In presence of a D-instanton 
half of the supersymmetries are broken, and there are sixteen  
supersymmetric zero modes or, equivalentely,
supersymmetric collective coordinates. To have a non vanishing
interaction between the  D-instanton and anti-D-instanton, the sixteen zero
modes must be saturated. Integrating over the fermionic collective coordinates
is equivalent to summing over all components of the supersymmetric multiplet
to which the D-instanton belongs. We can give a fully
 fledged string treatment in the context of the
boundary state formalism, for the case of D-instanton interactions in
flat space \refs{\greenal,\jose,\greengut}.
 In the low-energy approximation, the
D-instanton multiplet is replaced by a set of effective operators  
with up to sixteen fermionic legs. By the usual supersymmetric power-counting,
a pair of fermions corresponds to one derivative. Therefore, to leading
order in the low-energy expansion, 
an effective operator with $N_f$ fermions must have $8 -N_f/2$ 
derivatives to saturate the sixteen zero modes.

     We should think of these effective operators 
 as the result of having integrated out the massive
string modes. Therefore, for consistency, the instanton/anti-instanton
gases that we consider must be {\it dilute} in the sense of the
superstring background, i.e. the proper distance between topological
defects should be much larger than the string scale. In this way the
short-distance stringy singularities of the I/A interaction are also
avoided.  The restriction to dilute configurations is also technically
required  
 for the I/A pair to be well-defined as an
approximate non-perturbative configuration. This is important because
the I/A pairs are in the same  topological sector as the perturbative
excitations.

The full set of effective operators  in type IIB string theory, contributing 
up to relative order $O(\alpha'^3)$ can
be summarized in the $(R_{\mu\nu\rho\sigma})^4$
terms of the type IIB supergravity  together with their  
superpartners. Such operators are explicitly constructed for all 
channels in refs.\ \refs{\greengut}. Therefore, the supergravity 
approximation can
be extended to the analysis of I/A interactions in  curved backgrounds, such
as $AdS_5 \times {\bf S}^5$.

We can thus compute the D-instanton/anti-D-instanton interaction by
evaluating the vacuum amplitudes of these effective operators. Among
them, the most characteristic is the operator with sixteen dilatinos
and no derivatives (this operator  determines the instanton measure),
 for which the I/A interaction is simply given by the
single fermionic zero mode exchange.
The corresponding effective operator reads (in the string frame):
\eqn\sixteenop
{I_{\lambda^{16}} = {1\over \alpha'} \int d^{10} x \,\sqrt{-g} \;
e^{-\phi/2} \,f_{(16)} (\tau,{\bar \tau}) \;\varepsilon_{[16]} \; 
(\lambda^\alpha)^{16},}
where $\varepsilon_{[16]}$ denotes the completely antisymmetric tensor and
$f(\tau, {\bar \tau})$ is a modular function of the type IIB complex
coupling $\tau = ie^{-\phi} -\chi$. To leading order in the weak-coupling
expansion, the single-instanton contribution to $f_{(16)}$   is 
proportional
to $e^{-12\phi} \,e^{2\pi i\tau}$, where we recognize the typical 
instanton factor $e^{-2\pi/g_s} = e^{-8\pi^2 /g_{\rm YM}^2}$.    

 In the rest of the paper we will
concentrate mostly on this contribution to the I/A interaction action,
postponing for another publication a more detailed analysis of all
channels.           
The corresponding vacuum diagrams take the form 
\eqn\exch{
W_{\rm I/A} = ({\rm V.F.})\; \prod_{\rm Fermi \; lines} 
\left\langle I \,|\,\Dirac^{-1} \,|\,A \right\rangle.} 
with  instantons or anti-instantons contributing sixteen oriented
fermion lines each. The term (V.F.) represents the contribution of
vertex factors coming from \sixteenop, 
with the instanton action $e^{-2\pi/g_s}$ and
the coupling constant dependence of the instanton measure, both included
in $f_{(16)}$,  as well as a totally antisymmetric tensor  contracting
all spinor indices at each vertex. Explicitly, in Poincar\'e coordinates:
\eqn\vf{
({\rm V.F.}) = \prod_{\rm vertices} \sqrt{N} \; R^{8\cdot 9} \;N^{-16}\;\varepsilon_{[16]} \;
\int {d\rho \,d{\vec x} \over \rho^5} \,d\Omega_5 \;f_{(16)},}
where the  power of $R$ comes from the $\alpha'$ and $g_s$ 
dependence in \sixteenop\ and the rescaling $\lambda^\alpha \rightarrow
N^{-1} \,R^4 \lambda^\alpha$ to have canonically normalized dilatinos.

The chirality selection rules
forbid Fermi exchange interaction between I/I or A/A pairs, a fact that
we shall see explicitly bellow.   
We will be primarily interested in the spatial dependence of the interaction
action. Therefore,  we shall focus   
on the structure of the   Dirac propagator  
between  an instanton located at $\vec{x}$ and an
anti-instanton at $\vec{y}$,  ``far apart'', but with approximately the
same scale size $\rho_x \sim \rho_y$, and at the same point in the
five-sphere $\Omega_x =\Omega_y$. 
Our  main observation is the existence of two dynamical regimes for
I/A interactions, that is two main regimes of interest for 
the evaluation of the propagator. 

If the geodesic distance $d_{xy}$ is small compared to the AdS and 
sphere radius
$R\sim \sqrt{\alpha'} \lambda^{1/4}$,
but still large compared to the string length scale: 
$\ell_s=\sqrt{\alpha'} \ll d_{xy} \ll R$, 
then the propagation is locally equivalent
to the ten-dimensional fermion propagation in flat space, i.e.\ we have
the following scaling of the propagator: 
\eqn\tend{
S(x,y)= \left({1\over \Dirac}\right)_{xy} \simeq - 
\dirac \,\left({1\over -\partial^2}\right)_{xy} \simeq  
 -\dirac\,
\left({1 \over d_{xy}}\right)^8   
,}  
up to corrections of order $O(d_{xy}/R)$. On the other hand,   
the geodesic distance in terms of Poincar\'e coordinates
for $d_{xy} \ll R$ is  proportional to the so-called ``cordal distance'':  
\eqn\cordal
{(d_{xy})_{\rm AdS}^2 \simeq R^2 {|{\vec x}-{\vec y}|^2 +
 (\rho_x - \rho_y)^2 \over \rho_x
\rho_y}      
}
so that, for instantons of the same scale size, and located at the
same point in the ${\bf S}^5$ we  have, up to corrections again of
order $O(d_{xy}/R)$
\eqn\app
{S(x,y) \simeq -\dirac \;
\left({1\over  d_{xy}}\right)^8 
\sim R^{-9} \,\left({\rho \over |{\vec x}-{\vec y}|} \right)^9
\; {\vec \Gamma} \cdot {\vec u}_{xy}   
}
where ${\vec u}_{xy} \equiv ({\vec x} - {\vec y})/ |{\vec x} - {\vec y}|$.
Finally, putting all pieces together, we find    
\eqn\uu
{\left(W_{\rm I/A}\right)^{\rm overlapping}
 = ({\rm V.F.})\;\prod_{\rm Fermi \; lines}  \left({
\rho \over |{\vec x}-{\vec y}|} \right)^9 
\; {\cal P}_x \,{\vec \Gamma}\;{\cal P}_y \,\cdot {\vec u}_{xy}\ +  \dots   
}
The dots standing for corrections proportional to $\rho_x -\rho_y$ and $\Omega_{xy}$, in addition
to higher orders in the small $d_{xy} /R$ expansion. 
In particular, for a single I/A pair we have a total power of sixteen in
the previous expression. 
We have made explicit the chiral structure by the insertion of the
ten-dimensional chiral projectors ${\cal P}_x , {\cal P}_y$, since 
the instanton vertices are chiral. A Fermi line must connect
an I/A pair, rather than a I/I or A/A pair, in order for the action
to be non-vanishing, i.e. we have the selection rule: 
${\cal P}_x (1-{\cal P}_y) = {\cal P}_x $. 
We have also suppressed the explicit
power of $R$ coming from eq. \app. In fact, it cancels against the
explicit $R$-dependence of the vertex factors in   
\vf, since propagators and vertices are in relation eight to one. 

 Notice that in this regime $|{\vec x}-{\vec y}| \ll \rho$, which can be
interpreted as the fact that the size of the instantons is much larger
than their distance, and thus they strongly overlap. This is a situation
which cannot be studied in perturbation theory, but it is consistent in
the supergravity approximation, because the D-instanton system is still
dilute in $AdS_5 \times {\bf S}^5$
 for $d_{xy} \gg \ell_s$. The elementary transition amplitude in string
units  
\eqn\dimcomb{
\ell_s^9 \;S(x,y) \sim \lambda^{-9/4} \, \left({
\rho \over |{\vec x}-{\vec y}|} \right)^9
,}
is still small in the window 
$|{\vec x}-{\vec y}| \ll \rho \ll \lambda^{1/4} \;|{\vec x}-{\vec y}| $, 
which is wide for $\lambda \gg 1$. This is
entirely analogous to other situations in AdS/CFT for non-BPS quantities.
Namely, large renormalizations of physical scale by powers of $\lambda$ are frequent. Examples
include the renormalization of the topological susceptibility in the
models of \refs{\topsus}, and the different scales of finite size 
effects in thermal partition functions on the torus, studied in \refs{\kogan}.

In the other dynamical regime, the geodesic distance is much larger than 
the AdS and sphere radius: $d_{xy} \gg  R$.
In this case $|{\vec x}-{\vec y}| \gg \rho$ and
we are in the more standard (in field theory terms) dilute instanton
regime. Now the scale of propagation is sensitive to the curvature of
the background. In practice, we have just to evaluate the fermionic 
propagator in the bulk of $AdS_5$. Indeed, since higher harmonics 
of ${\bf S}^5$ have a large Kaluza--Klein mass of order 
$M_{\rm KK} \sim 1/R$, their contribution is exponentially 
suppressed for propagation over distances $d_{xy} \gg R$. 
Moreover, the dilatini component along ${\bf S}^5$ is given by a 
Killing spinor, and the net effect of the sphere is  
summarized in a dimensional factor of the volume:
\eqn\dimanal{
 S(x,y)_{\rm 10d} \simeq R^{-5} \; S(x,y)_{
\rm AdS}\ . 
}
Upon Kaluza--Klein reduction on the ${\bf S}^5$, the dilatini acquire
an effective mass $m_f = -3/(2R)$, c.f. \refs{\kkred}. 
The fermionic propagator in the bulk of
$AdS_5$ has been already discussed in the literature \refs{\burg,\diracprop}, 
but it can be easily obtained as follows.
In Poincar\'e coordinates, it holds~\foot{We rescale $R$ out of all the
following equations, setting also $m_f=-3/2$.}
\eqn\fprops{
\left(\gamma^\mu D_\mu\right)^2 = \nabla^2 + 4 + \rho\, \gamma^5\, 
\vec{\gamma}\cdot \vec\partial, \qquad\qquad\qquad 
\left[\gamma^\mu D_\mu, \gamma^5\right] = 2 \rho\,
\vec{\gamma} \,\gamma^5\, \vec\partial
,}
where
\eqn\dslash{
\gamma^\mu D_\mu = \rho \, \gamma^5 \partial_\rho + \rho\, 
\vec{\gamma}\cdot\vec\partial -2\gamma^5
.}
$\gamma^\mu$ are the five-dimensional Dirac matrices, $\gamma^5$ refers to
the $\rho$ coordinate and $\nabla^2$ is the five-dimensional
 scalar laplacian. 
{}From these equations one obtains that the Dirac propagator for a
fermion with mass $m_f$ satisfying 
$(\gamma^\mu D_\mu -m_f)S = \delta^{(5)}$ is given by
\eqn\dprop{
S(x,y)= -\sqrt{{\rho_x\over \rho_y}} 
\left[\gamma_\mu D^\mu_y + {1\over2} \gamma^5 +
m_f \right] \left\langle x\,\Bigg| \, \left(\nabla_y^2 - \left[\left(
 {1\over2}
\gamma^5 + m_f\right)^2 -4\right] \right)^{-1}  \,\Bigg| y\right\rangle\ .
}
Thus the Dirac propagator for the component of a fermion, with well-defined
eigenvalue of $\gamma^5$ and mass $m_f$,  is related to
the propagator of a scalar with an effective mass
\eqn\effmass{
m^2_{\rm eff} = \left({1\over2} \gamma^5 + m_f\right)^2 -4\ .
}
The bulk bosonic propagator for a scalar is well known \refs{\bprop, \burg}, 
and given by
\eqn\bosprop{
G^B_\Delta (u) = {\Gamma(\Delta) \Gamma(\Delta -3/2) \over (4\pi)^{5/2}
\Gamma(2\Delta -3)} \, u^\Delta \, F(\Delta, \Delta -3/2, 2\Delta -3, -u) 
,}
where $u = 4 \rho_x\, \rho_y / (|{\vec x}-{\vec y}|^2 + (\rho_x -
\rho_y)^2)$ and using eq.\ \effmass\ 
\eqn\dDelta{
\Delta = 2 + \left| { 1\over 2} \gamma^5 + m_f \right|\ .
}
It is then easy to see that for instantons of the same scale size but 
large separation, i.e.\
large $|\vec x - \vec y|$, the fermionic propagator behaves as
\eqn\propdil{
{\cal P}_{\pm} \, S(x,y) \simeq c_1 {\cal P}_\pm \,\left(m_f + 
\left(\Delta_\pm -
{3\over 2}\right)\,\gamma^5 \right)\cdot
\left({\rho \over | \vec y - \vec x |}\right)^{2\Delta_\pm}
 + c_2 {\cal P}_\pm \;\vec{\gamma}\cdot {\vec u}_{xy}\;
\left({\rho \over | \vec y - \vec x |}\right)^{2\Delta_\pm +1} 
}
where the matrices ${\cal P}_\pm = {1\over 2} (1\pm \gamma^5)$ project 
onto definite $\pm 1$ eigenvalues of
$\gamma^5$, which is interpreted as the chirality eigenvalue in 
the four-dimensional boundary, i.e.
the gauge-theory space-time.  $\Delta_\pm$  is given by $\Delta$  
in eq. \dDelta\
above with $\gamma^5 = \pm 1$.  In our case $m_f = -3/2$ and the terms with $\Delta_+ =3$ 
dominate over those with
$\Delta_- = 4$ in the limit of large $  | \vec y - \vec x |/\rho$.
   Of these, the naively leading one is 
proportional to $c_1 {\cal P}_+ 
{\cal P}_- =0$ and thus vanishes. The actual leading term is chiral 
in four-dimensional terms and 
leads to   a net I/A interaction of the form  
\eqn\dilutescal{
\left(W_{\rm I/A} \right)^{\rm dilute} = ({\rm V.F.})\;
\prod_{\rm Fermi \; lines}
 \left({\rho\over | \vec y - \vec x |} \right)^7
\;  {\cal P}_+\,\vec{\gamma}\cdot {\vec u}_{xy} \ +\, \dots}
with the dots representing higher corrections in powers of $\rho_x -\rho_y$ and integer powers
of $u$. Thus, we can distinguish the dilute and overlapping regimes in this
channel by the overall power of the physical separation in the gauge
theory $| \vec y - \vec x |$, provided $\lambda $ is large enough to
justify the various approximations.
Notice that for a A/I pair the four-dimensional chirality
of the amplitude turns out to be opposite, i.e.\ ${\cal P}_-$ appears in 
$W_{\rm A/I}$, as one expects from the action of a  CP transformation.

\newsec{The Non-Commutative case}

In the recent works \refs{\hi} and \refs{\mr} a gravitational dual of
large $N$ NCYM was proposed as the near-horizon geometry of D$p$-branes 
\refs{\ncdp, \ncyaron, \ncthermobr}
with non-trival profiles of NS $B$-fields, under the 
appropriate scaling of parameters that isolates the low energy regime
with NCYM physics. Let us consider
the particular case of an euclidean  D3-brane system
endowed with a constant $B$-field
background of rank two with skew-eigenvalues $B_z, B_w$ in the respective
planes $z= x_1 + ix_2, \; w= x_3 +ix_4$.  We have   
the  string-frame geometry: 
\eqn\ncdp{
{ds^2 \over R^2} = {1\over \rho^2}  \left( {\hat f}_z (\rho) \,|dz|^2 
 + {\hat f}_w (\rho)\, |dw|^2 \right) + 
{d\rho^2 \over \rho^2}  + d\Omega_{5}^2 
}
in terms of the functions 
\eqn\ncf{
{\hat f}_z   (\rho) = {\rho^4 \over \rho^4 + \rho_z^4}
,\qquad {\hat f}_w   (\rho) = {\rho^4 \over \rho^4 + \rho_w^4}
\ . 
}
The relevant  length scales introduced in the radial profile
by the noncommutativity properties are  related to the perturbative
non-commutative  length squared $[z,{\bar z}] = -2i\, \theta_z, \; 
[w,{\bar w }] = -2i\, \theta_w $ by the formulas 
\eqn\delta{
\rho_z = (\lambda \, \theta^2_z )^{1/4},\qquad
\rho_w = (\lambda \, \theta^2_w )^{1/4}
\ .
}
In these expressions, $\lambda = g_{\rm YM}^2 \,N$ is the 't Hooft
coupling of the Yang--Mills theories normalized in the large-$\rho$
region --- that is in terms of the ordinary commutative theory that 
appears in the infrared.
The dilaton and NS $B$-field have the profiles
\eqn\dilp{ 
e^{2\phi} = {\lambda^2 \over 16\pi^2 N^2} \,
{\hat f}_z (\rho) \;{\hat f}_w (\rho)\ ,\qquad
B_z =  {s_z \over \theta_z}\, {\rho_z^4 \over \rho^4 + \rho_z^4}
,\qquad B_w =  {s_w \over \theta_w}\, {\rho_w^4 \over \rho^4 + \rho_w^4}
\ .
}
Here, $s_z$ and $ s_w$ are sign factors controlling the sign 
$S_{\rm Pf} = s_z \, s_w$ of the pfaffian  ${\rm Pf}(B) = B_z B_w$ 
(we can choose, without loss of generality $\theta_i \geq 0$). 

Ramond--Ramond (RR) fields coupling to any non-trivial 
product $B\wedge B\wedge \cdots\wedge B$ on the D$p$-brane world-volume 
are also excited.   
For the case of interest here, the type IIB two- and zero-form
RR field strengths are excited. In particular, the axion profile is given
by \refs{\mr} 
\eqn\axpro{
\chi = -{\theta_{\rm YM} \over 2\pi} + 
i\,{4\pi N \over \lambda} \, S_{\rm Pf} \;  {\rho_z^2 \rho_w^2
 \over \rho^4},}      
where $\theta_{\rm YM}$ is the Yang--Mills vacuum angle, also normalized
in the infrared.

Since ${\hat f} (\rho\rightarrow 0)\rightarrow 0$, there is a significative
distortion of the metric and various field profiles \ncdp, \dilp\ in  
the ultraviolet regime. This fact has consequences for the interpretation
of holography in these dual large $N$ descriptions of the NCYM theory.
For example, there are problems in defining correlators of local
operators \refs\mr\ related to ambiguities in the ultraviolet 
renormalization.

A case that exposes these difficulties in a dramatic way is that of euclidean
D3-branes with a (anti-) self-dual   antisymmetric tensor 
$B^+ =0 \; (B^- =0)$ or   
$\theta_z =\theta_w = \theta,  \; \rho_z = \rho_w = \rho_\theta$ in the   
previous formulas. The Einstein frame metric, appropriate to the discussion
of massless field propagation,  is given by
\eqn\emet{
ds_E^2 = \left({r_\theta \over R}\right)^2 \,{1\over \sqrt{1+(r_\theta / 
r)^4}} \; d{\vec x}^{\,2} + \left({R\over r_\theta}\right)^2 \, \sqrt{
1+(r_\theta /r)^4} \; \left( dr^2 + r^2 \, d\Omega_5^2 \right)\ . 
}
With the identifications $r= R^2 /\rho$, $r_\theta = R^2 / \rho_\theta$
and certain rescalings of the coordinates,
this is simply the full D3-brane metric asymptotic to flat space 
as $r\rightarrow \infty$   (the
other fields do not correspond however to this solution). Since the
resulting manifold is asymptotic to flat ${\bf R}^{10}$ it is not
obvious in what way this description can be holographic. The crossover
from the ``throat" to the flat region is at $\rho = \rho_\theta$, i.e.\
at the onset of non-commutative effects in the gauge theory.

Despite these problems in making sense of local operators in the
geometric picture, one can easily compute certain observables, such
as the thermodynamic quantities, that can be written as integrals
over spacetime of local operators \refs{\ncthermobr, \ncthermoo}.

Non-commutative instantons provide an interesting probe into the locality
properties of NCYM theories \refs{\nekrasov,\sw}.
It is natural to try to draw some lesson from the AdS/CFT picture of these
in terms of D-instantons in the geometry \emet.    Some general properties
expected for instantons can be obtained from the Dirac--Born--Infeld action
of a D-instanton probe: 
\eqn\dbi{
S_{\rm DBI} = 2\pi \left(e^{-\phi} -i\chi\right)\ , 
}
with anti-D-instantons coupling  instead to the conjugate
combination of type IIB  dilaton and axion:   $e^{-\phi } +i\chi$. 
Upon substitution of the previous formulas we find
\eqn\probep{
S_{\rm DBI} = {8\pi^2 \over g_{\rm YM}^2} 
\left[ \sqrt{\left(1+{\rho_z^4 \over  
\rho^4} \right) \left(1+ {\rho_w^4 \over  \rho^4}
 \right)} + S_{\rm Pf} \; {\rho_z^2
\rho_w^2 \over \rho^4}\right] +i\,\theta_{\rm YM}\ .
} 
We see that, for very large instantons $\rho \gg \rho_z, \rho_w$ compared
to the non-commutativity scales, the action reduces to the usual 
$i\,\theta_{\rm YM} + 8\pi^2 /g_{\rm YM}^2$. On the other hand, for 
$\rho \sim \rho_z, \rho_w$ and $\rho \ll \rho_z, \rho_w$ there are 
significative modifications. In fact, the radial coordinate $\rho$ is
not an exact moduli for D-instantons  in the generic case. In other words,
the D-instanton probe is to be considered an approximate or constrained
instanton. For $S_{\rm Pf}=1$ the large action for small ``sizes'' 
could be interpreted as the absence of a small instanton sigularity in 
instanton moduli space. 
In particular, for $B^- =0$ we have $S_{\rm Pf} = 1$ and 
\eqn\dbsup{
S_{B^- =0} = {8\pi^2 \over g_{\rm YM}^2} + i\theta_{\rm YM} + 
{16\pi^2 \over g_{\rm YM}^2} \; \left({\rho_\theta \over\rho}\right)^4 
,} 
giving a strong suppression of the ``small'' instanton regime. 
On the other hand, for $B^+ =0$ one has $S_{\rm Pf} = -1$ and 
\eqn\dbjust{
S_{B^+ =0} = {8\pi^2 \over g_{\rm YM}^2} +i\,\theta_{
\rm YM}
}
exactly as in YM theory. Namely, non-commutative instantons in  a self-dual
$B$-field (conversely anti-instantons in an anti-self-dual $B$-field) have
a standard commutative moduli space. In particular there is an apparent
small instanton singularity. It is very satisfying to see these results
emerge in such an elementary way from the AdS/CFT picture in terms of
D-instanton probes. 

On the other hand, even if the moduli space of instantons   only depends    
on the self-dual part of the non-commutative deformation parameter, in
general one expects the precise form of the instanton solutions to depend
also on $B^-$. Therefore, it is not clear to what extent the region
$\rho \ll \rho_\theta$ really represents small non-commutative instantons,
in a physical sense.

A preliminary test of the locality properties of such $\rho \ll \rho_\theta$
or ``small" instantons can be extracted from our previous analysis of 
the I/A interaction.  This is a particularly interesting quantity
in the case at hand, since it does not involve directly the specification
of boundary behaviour at $\rho=0$, and thus  it should be free of
renormalization ambiguities.

The D-instanton probe description of non-commutative
instantons is generically off-shell for $\rho\ll \rho_\theta$.
For example,  let $B^+ =0$ and place a Dirichlet-(I/A)  
pair  at large $\rho$. From eq.\ \dbjust\ it follows that
we can ``drag'' the D-instanton to $\rho \ll \rho_\theta$ with no cost
in action, but from eq.\ \dbsup\ it follows that it is not possible, 
without a large cost in action, to drag the corresponding
anti-D-instanton to $\rho \ll \rho_\theta$.

If we insist in mantaining the off-shell Dirichlet-(I/A) pair 
at ``small" size (i.e.\ $\rho\ll \rho_\theta$), as an approximate
or ``constrained" configuration,  then  
the fermionic exchange interaction over distances 
 $d_{xy} \gg \rho_\theta$ 
must be evaluated in  the geometry \emet, which is asymptotic to flat 
${\bf R}^{10}$ in the region of interest ($\rho \ll \rho_\theta$) for
both ``instantons''. Therefore, the appropriate fermion propagator
shows ten-dimensional scaling
\eqn\ncsca{
S(x,y) =  \left( {1\over \Dirac 
}\right)_{xy} \simeq 
-\dirac \, \left( {1\over d_{xy}}\right)^8   
\sim R^{-9} \; \left({\rho_\theta \over |{\vec y} -{\vec x}|}\right)^9 \,
\;{\vec \Gamma}\cdot{\vec u}_{xy}  
,}
where we have used the geodesic distance computed in the metric \emet. We 
find an effective interaction  of the overlapping type, see eq.\ \app,    
with a characteristic scale independent of the size $\rho$ and fixed 
at the non-commutative length $\rho_\theta\gg\rho$. 

A possible interpretation of this result is that, as the 
instantons decrease in size to the scale where non-commutative effects
start to be relevant, they actually ``delocalize", behaving again as if 
they were large, but with an effective size dictated by the non-commutative
geometry.

Finally, when  considering the interaction of a D-instanton and an
anti-D-instanton of large size (i.e.\ $\rho \gg \rho_\theta$), the
results of the previous section apply since the
metric \emet\ in this region is asymptotically given by 
$AdS_5 \times {\bf S}^5$.


\newsec{Acknowledgements}
This work is partially supported by the European Commission TMR programme
ERBFMRX-CT96-0045 in which A.P.\ is associated to the Milano University.
A.P.\ would like to thank CERN and the CIT-USC Center for Theoretical
Physics for their hospitality while part of this work was carried out.

\listrefs

\bye